\documentclass[sigconf]{acmart}
\usepackage{algorithm}
\usepackage{algorithmic}
\usepackage{balance}

\AtBeginDocument{%
  }

\copyrightyear{2025}
\acmYear{2025}
\setcopyright{acmlicensed}\acmConference[WWW Companion '25]{Companion Proceedings of the ACM Web Conference 2025}{April 28-May 2, 2025}{Sydney, NSW, Australia}
\acmBooktitle{Companion Proceedings of the ACM Web Conference 2025 (WWW Companion '25), April 28-May 2, 2025, Sydney, NSW, Australia}
\acmDOI{10.1145/3701716.3715576}
\acmISBN{979-8-4007-1331-6/2025/04}

\begin{document}

\title{Scalable Overload-Aware Graph-Based Index Construction for 10-Billion-Scale Vector Similarity Search}

\author{Yang Shi}
\email{shiyang1@xiaohongshu.com}
\affiliation{%
  \institution{Xiaohongshu Inc}
  \city{Shanghai}
  \country{China}
}

\author{Yiping Sun}
\authornote{Corresponding Author.}
\email{sunyiping@xiaohongshu.com}
\affiliation{%
  \institution{Xiaohongshu Inc}
  \city{Shanghai}
  \country{China}
}

\author{Jiaolong Du}
\email{jiaolong@xiaohongshu.com}
\affiliation{%
  \institution{Xiaohongshu Inc}
  \city{Shanghai}
  \country{China}
}

\author{Xiaocheng Zhong}
\email{mingcheng1@xiaohongshu.com}
\affiliation{%
  \institution{Xiaohongshu Inc}
  \city{Shanghai}
  \country{China}
}

\author{Zhiyong Wang}
\email{sunzhenghuai@xiaohongshu.com}
\affiliation{%
  \institution{Xiaohongshu Inc}
  \city{Shanghai}
  \country{China}
}

\author{Yao Hu}
\email{xiahou@xiaohongshu.com}
\affiliation{%
  \institution{Xiaohongshu Inc}
  \city{Shanghai}
  \country{China}
}

\renewcommand{\shortauthors}{Yang Shi et al.}

\begin{abstract}
  Approximate Nearest Neighbor Search (ANNS) is essential for modern data-driven applications that require efficient retrieval of top-k results from massive vector databases. Although existing graph-based ANNS algorithms achieve a high recall rate on billion-scale datasets, their slow construction speed and limited scalability hinder their applicability to large-scale industrial scenarios. In this paper, we introduce \textbf{SOGAIC}, the first \textbf{S}calable \textbf{O}verload-Aware \textbf{G}raph-Based \textbf{A}NNS \textbf{I}ndex \textbf{C}onstruction system tailored for ultra-large-scale vector databases: 1) We propose a dynamic data partitioning algorithm with overload constraints that adaptively introduces overlaps among subsets; 2) To enable efficient distributed subgraph construction, we employ a load-balancing task scheduling framework combined with an agglomerative merging strategy; 3) Extensive experiments on various datasets demonstrate a reduction of $47.3\%$ in average construction time compared to existing methods. The proposed method has also been successfully deployed in a real-world industrial search engine, managing over 10 billion daily updated vectors and serving hundreds of millions of users.
\end{abstract}



\maketitle

\section{Introduction}
Approximate Nearest Neighbor Search (ANNS) has been extensively studied in many recent research \cite{ootomo2024cagra,sun2024real,khan2024bang} due to its crucial role in a wide range of applications, such as web search engines, recommendation systems, and Retrieval-Augmented Generation (RAG) \cite{jiang2024piperag} for large language models (LLMs). These applications rely heavily on ANNS for efficient retrieval of top-k results. Among all ANNS algorithms, graph-based ones \cite{wang2021comprehensive} such as HNSW\cite{malkov2018efficient}, NSG\cite{fu2017fast}, and NGT\cite{iwasaki2018optimization} have emerged as highly effective paradigms due to their ability to represent neighbor relationships in a graph structure, which significantly reduces the computational burden required for high-quality retrieval, especially comparing to space-partitioning methods such as IVF, KD-tree \cite{ram2019revisiting}, and LSH \cite{huang2015query}. Recently, approaches like DiskANN \cite{jayaram2019diskann} have been developed to construct SSD-resident ANNS graph indexes for billion-scale search, minimizing disk I/O while using limited main memory. However, their index construction process, based on divide-and-conquer \cite{bentley1980multidimensional}, is slow and hard to scale, hindering daily or more frequent updates of vector embeddings in real-world applications. The difficulty arises from two key challenges:

Firstly, efficiently creating overlapped subset divisions for large datasets with overload-aware considerations remains unexplored. DiskANN introduces overlap among subsets by assigning each vector to a fixed number of subsets based on its distance to centroids from clustering. However, this approach often results in redundant overlap between geometrically close divisions and sometimes fails to ensure balanced vector assignments due to the uneven distribution of points in the clusters, a consequence of K-means clustering with limited samples. This imbalance can cause a subset's subgraph construction to exceed resource limits, also known as overload issues, influencing the overall process. While density-based methods like DBSCAN offer more precise divisions, they are impractical for large datasets due to high computational costs, parameter sensitivity, and centroid control challenges. Thus, a more adaptive data partitioning approach is needed to reduce redundant assignments while keeping the overload boundary in check.

Secondly, an efficient and scalable scheduling framework for building and merging ANNS subgraphs remains unaddressed. Existing methods, such as DiskANN, often rely on a single high-cost machine to sequentially build and merge subgraphs, limiting scalability and efficiency. To address this, SPTAG \cite{ChenW18} employs a scalable k-NN graph construction method \cite{WangWZTGL12} to partition datasets into non-overlapping subsets and distribute tasks across a cluster. However, as datasets grow in size and complexity, optimizing subgraph overlap for connectivity requires extensive iterations, leading to redundant computations and storage overhead. Observing that subgraph construction time is proportional to the size of its subset, an efficient scheduling framework can be designed to minimize construction time on a cluster without additional redundant computations, provided that overlap coverage and overload constraints are carefully addressed by a cost-effective data partitioning approach.

To address these challenges, we introduce \textbf{SOGAIC}, the first \textbf{S}calable \textbf{O}verload-Aware \textbf{G}raph-Based \textbf{A}NNS \textbf{I}ndex \textbf{C}onstruction system, designed for ultra-large-scale vector databases exceeding 10 billion points. \textbf{SOGAIC} accelerates index construction through an adaptive data partitioning strategy coupled with a load-balancing task scheduling framework, ensuring scalability with computational resources while maintaining high-quality graph structures for efficient vector similarity search.

\section{System Overview}

\begin{figure}[t]
  \includegraphics[width=0.95\columnwidth]{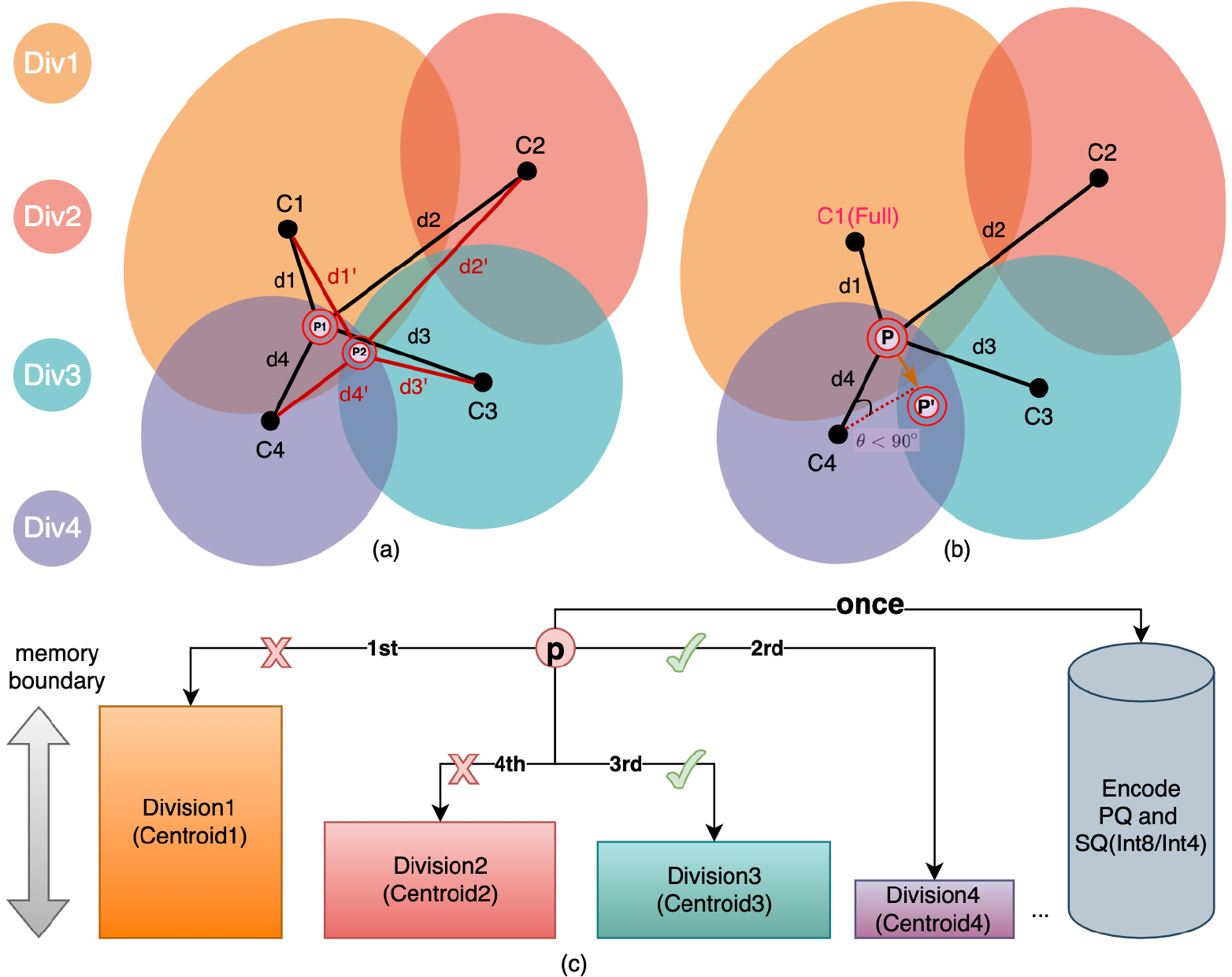}
  \caption{(a) Illustration of the assignment process for \textbf{P1} and \textbf{P2} based on their geometric relationships with centroids \{\textbf{C1} ... \textbf{C4}\}. (b) Illustration of the deviation from \textbf{P} to \textbf{P'} for the assignment process of \textbf{P}, when overload happens on division1. (c) Illustration of the overall data partitioning process for \textbf{P} with quantization encoding executed in parallel.}
  \Description{Three diagrams}
  \label{Figure 1}
\end{figure}

The proposed system consists of two main stages. The first stage handles overlapped subsets division using an adaptive data partitioning algorithm while controlling overload bounds. The second stage schedules subgraph construction tasks across a distributed cluster and executes the agglomerative merging strategy to form the final ANNS graph index.

\begin{algorithm}[t]
    \caption{Adaptive Overload-Aware Vector Assignment}
    \label{alg:partition}
    \begin{algorithmic}[1]
        \REQUIRE Vector to be assigned $\boldsymbol{\mathcal{V}}$
        \REQUIRE Centroids from K-means $c_i \in \boldsymbol{\mathcal{C}}$
        \REQUIRE Max overlapping factor $\boldsymbol{\Omega}$ >= 2
        \REQUIRE Adaptive relaxation parameter \( \epsilon \) > 1
        \REQUIRE Initialized sets for all centroids $s_i \in \boldsymbol{\mathcal{S}}$
        \REQUIRE Maximum assignable vectors of a set $\boldsymbol{\Gamma}$
        
        \STATE \textbf{Initialize:} Create a priority queue \textbf{\textit{Q}<CentroidIndex, Dist>}
        \FOR{$i = 0$ \TO $|\boldsymbol{\mathcal{C}}| - 1$}
            \STATE Insert $(i, \text{distance}(\boldsymbol{\mathcal{V}}, c_i))$ into $\textbf{\textit{Q}}$
        \ENDFOR
        
        \STATE \textbf{Initialize:} $curOLPCnt \leftarrow 0$; $curOLPFactor \leftarrow 0$
        \STATE \textbf{Initialize:} $accDist \leftarrow 0$; $curAVGDist \leftarrow \infty$
        \STATE \textbf{while} \textbf{\textit{Q}} is not empty \textbf{and} curOLPCnt < $\boldsymbol{\Omega}$
        \STATE \hspace{8pt} $(\textbf{index}, \textbf{dist}) \leftarrow \textbf{ExtractMin}({\textbf{\textit{Q}}})$ 
        \STATE \hspace{8pt} \textbf{if} \textbf{dist} <= \( \epsilon \) $*$ curAVGDist
        \STATE \hspace{16pt} $\textit{curOLPFactor} \leftarrow \textit{curOLPFactor} + 1$
        \STATE \hspace{16pt} $\textit{accDist} \leftarrow \textit{accDist} + \textit{dist}$
        \STATE \hspace{16pt} $\textit{curAVGDist} \leftarrow \textit{accDist} / \textit{curOLPFactor}$
        \STATE \hspace{16pt} \textbf{if} |$s_{index}$| \textbf{<} $\boldsymbol{\Gamma}$
        \STATE \hspace{24pt} $\textit{curOLPCnt} \leftarrow \textit{curOLPCnt} + 1$
        \STATE \hspace{24pt} $s_{index} \leftarrow s_{index} \cup \{ \boldsymbol{\mathcal{V}} \}$
        \STATE \hspace{16pt} \textbf{else}
        \STATE \hspace{24pt} $curAVGDist \leftarrow \infty$
        \STATE \hspace{16pt} \textbf{end if}
        \STATE \hspace{8pt} \textbf{end if}
        \STATE \textbf{end while}
    \end{algorithmic}
\end{algorithm}

\subsection{Overload-Aware Adaptive Data Partitioning}

The algorithm for partitioning the vector database into multiple overlapping subsets begins by determining the maximum number of base points per division, referred to as \(\boldsymbol{\Gamma}\) (capacity), which is constrained by the memory limits of each container in the computational cluster. Given a dataset containing \(\boldsymbol{N}\) (total points) and a maximum overlapping factor \(\boldsymbol{\Omega}\) (defining the maximum number of partitions a point can belong to), we can estimate the minimum number of partitions \(\boldsymbol{\Phi}\) (number of centroids) required to handle highly imbalanced data distributions effectively, which is:
$\boldsymbol{\Phi} = \lceil \boldsymbol{\Omega} \times \boldsymbol{N} / \boldsymbol{\Gamma} \rceil$. 
After estimating \(\boldsymbol{\Phi}\), we apply K-means clustering on a small sample of the dataset to obtain \(\boldsymbol{\Phi}\) centroids $\boldsymbol{\mathcal{C}} = \{c_1, c_2, c_3, \dots, c_{\boldsymbol{\Phi}}\}$ as the initial reference points for vector assignment.

Instead of assigning each point to a fixed number of divisions, we propose a novel vector assignment method (Algorithm~\ref{alg:partition}) that allocates points to divisions based on their geometric relationships with centroids, while simultaneously constraining the overload bounds for each subset. Each point is assigned to a division if its distance to the centroid is less than the average distance (lines 10-12) of previous assignments, scaled by an adaptive relaxation parameter \( \epsilon > 1 \) (line 9). This parameter adapts to different data distributions: It should remain small for uniformly distributed datasets to minimize redundant assignments, while being larger for structured datasets that are challenging to capture using distance-based clustering methods like K-means. A larger parameter allows points to be assigned to more divisions, even when distances to some centroids are greater, fostering overlap between divisions and enabling the formation of long bridging edges among disconnected ANNS graphs. This improves graph connectivity, which is crucial for an efficient ANNS search with a high recall rate. Upon reaching the overload limit (line 13), the current average distance is reset (line 17) to ensure that each point is assigned to at least one division.

An example for Algorithm~\ref{alg:partition} is illustrated by Figure~\ref{Figure 1}(a) and Figure~\ref{Figure 1}(b). In Figure~\ref{Figure 1}(a), assuming no overload happens and $\boldsymbol{\Omega} = 3, \epsilon = 1.5$, the first assignment for \textbf{P1} is \( \textbf{C1} \leftarrow \textbf{C1} \cup \{ \textbf{P1} \} \) (line 15). Due to \( d_4 < 1.5 \times d_1 \) and \( d_3 > 1.5 \times \frac{\sum(d_1, d_4)}{2} \) (line 9-12), the final assignment is \( \textbf{P1} \in \textbf{C1} \cap \textbf{C4} \). For \textbf{P2}, since \( \textbf{C4} \leftarrow \textbf{C4} \cup \{ \textbf{P2} \} \), \( \textbf{C3} \leftarrow \textbf{C3} \cup \{ \textbf{P2} \} \), \( d_1' < 1.5 \times \frac{\sum(d_4', d_3')}{2} \), and $\boldsymbol{\Omega}$ is reached (line 7), the final assignment is \( \textbf{P2} \in \textbf{C4} \cap \textbf{C3} \cap \textbf{C1} \). In Figure~\ref{Figure 1}(b), assuming \textbf{C1} is full and $\boldsymbol{\Omega} = 3, \epsilon = 1.8$, since \textbf{C4} is the next nearest one (line 8), \( d_3 < 1.8 \times \frac{\sum(d_1, d_4)}{2} \), and \( d_2 > 1.8 \times \frac{\sum(d_1, d_4, d_3)}{3} \), the final assignment is \( \textbf{P} \in \textbf{C4} \cap \textbf{C3} \). This assignment diverges from the original geometric relationships between \(\mathbf{P}\) and the centroids \(\mathbf{C1}\) to \(\mathbf{C4}\). It can be interpreted as relocating \(\mathbf{P}\) from \(\mathbf{C1} \cap \mathbf{C4}\) to \(\mathbf{C4} \cap \mathbf{C3}\), resulting in a new point \(\mathbf{P'}\). However, with an appropriately chosen \(\epsilon\), this deviation can be constrained within an angle \(\theta < 90^\circ\) centered at \(\mathbf{C4}\), which has minimal impact on the neighborhood relationships. Our experiments, as well as the findings from the ANGN algorithm in \cite{pourbahrami2018survey}, demonstrate that the graph quality for ANNS is preserved under these conditions.

Finally, since the vector assignment process is independent and requires limited memory per task, the proposed system parallelizes the quantization encoding alongside the vector assignment without any redundant computations, offering greater efficiency than the sequential approach used by DiskANN. The quantization is usually used to accelerate distance approximations while reducing storage for similarity search. As shown in Figure~\ref{Figure 1}(c), each vector is encoded only once, and the resulting codes are merged in the subsequent stage, ensuring scalability and efficiency.

\subsection{Distributed ANNS Graph Construction with Agglomerative Subgraph Merging}

\begin{figure}[t]
  \includegraphics[width=1\columnwidth]{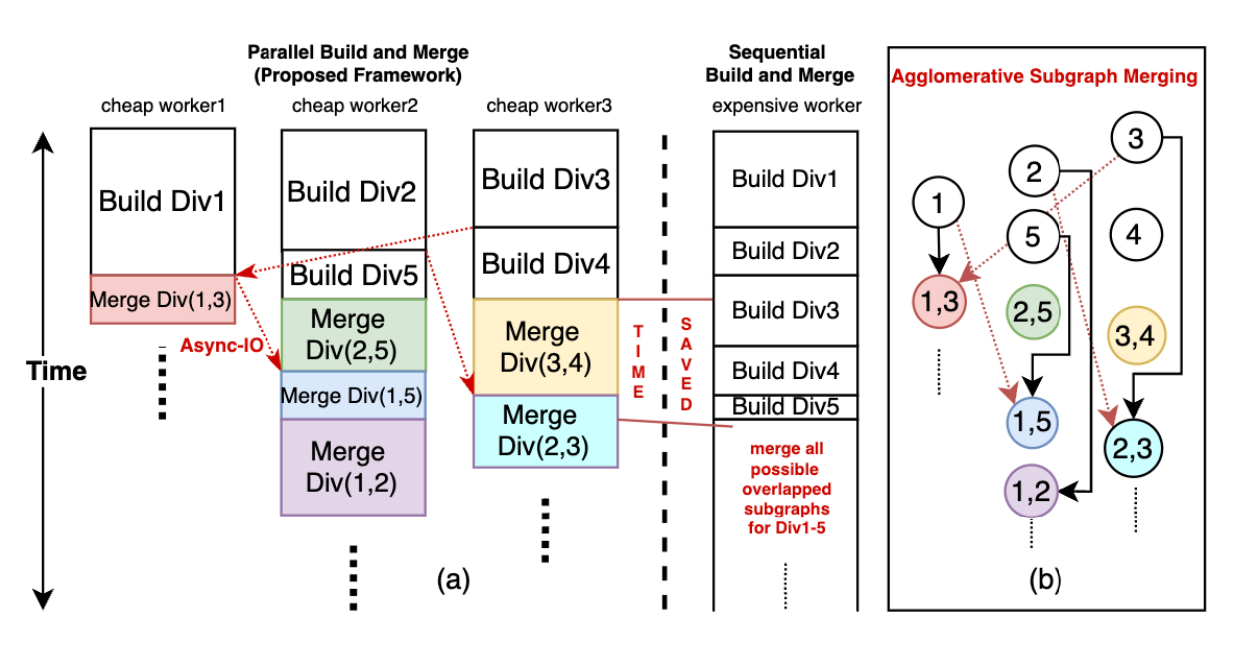}
  \caption{(a) Parallel build and merge framework versus Sequential one. (b) Agglomerative subgraph merging topology.}
  \Description{Three diagrams: (a) shows the geometric relationship of P1 and P2, (b) shows the geometric relationship of P, while }
  \label{Figure 2}
\end{figure}

This section introduces a comprehensive and highly efficient framework designed specifically for the construction and seamless merging of subgraphs within a distributed computing environment. First, we employ a load-balancing scheduling method for subgraph construction, leveraging the near-linear relationship between ANNS graph construction time and dataset size. As shown in Figure~\ref{Figure 2}(a), subsets produced by the previous stage (Section 2.1) are first sorted by size in descending order to prioritize larger tasks, which have a greater impact on load balancing. After sorting, graph-building tasks are iteratively assigned to the least-loaded machine or container, ensuring that each task contributes the least to the maximum load. By always choosing the least-loaded machine, the scheduling process greedily maintains a minimum overall load balance across all machines or containers in a cluster. Additionally, the maximum subset size is constrained by the previous stage (Section 2.1), mitigating overload issues such as out-of-memory errors and insufficient disk capacity. Otherwise, more advanced scheduling methods like BDSC \cite{khaldi2015parallelizing} or LSSP\cite{radulescu2002low} would be required. As a result, the system scales efficiently by incorporating low-resource workers to accelerate construction, rather than relying on a single high-resource worker.

Then, the agglomerative subgraph merging strategy, illustrated in Figure~\ref{Figure 2}(b), enables efficient merging by allowing immediate processing of completed subgraphs. Using distributed computing frameworks such as Apache Spark or MapReduce\cite{dean2008mapreduce}, subgraphs are asynchronously exchanged among different machines or containers to enable the next level of merging. The computationally intensive part involves neighbor selection for overlapping regions, while disjoint parts carry over to the next stage without additional computation. By employing a hierarchical or tree-based structure, the complexity decreases from \(O(n)\) in traditional methods to \(O(\log n)\). Unlike on-disk merging used in methods such as DiskANN, which must handle memory constraints when merging all subgraphs simultaneously, our approach prioritizes in-memory merging for subgraphs that fit within the main memory at each stage of execution. This strategy not only mitigates I/O bottlenecks but also enhances graph quality by allowing immediate access to vectors and neighbor candidate distances, leading to more precise pruning and selection. Furthermore, we optimize task scheduling by dynamically monitoring subgraph overlap counts, ensuring that merges with higher overlap receive higher priority.

\section{Experiment}

\subsection{Experimental Setting}
\setlength{\parindent}{0pt}\underline{\textbf{Datasets.}}
We conduct experiments on five datasets, including image embeddings (\textbf{\textit{SIFT1M}}, \textbf{\textit{SIFT1B}}, and \textbf{\textit{ISD3B}} - Our image search dataset), text embeddings (\textbf{\textit{GloVe}}), and video embeddings (\textbf{\textit{VDD10B}} - Our video de-duplication dataset). Their main characteristics are summarized in Table ~\ref{tab:datasets}. LID \cite{amsaleg2015estimating} indicates local intrinsic dimensionality which implies the hardness of a dataset.

\setlength{\parindent}{0pt}\underline{\textbf{Compared approaches.}} We selected the following methods for comparison with the proposed approach (\textbf{SOGAIC}): HNSW \cite{malkov2018efficient} as implemented in Faiss \cite{douze2024faiss}, DiskANN \cite{jayaram2019diskann}, and SPTAG \cite{ChenW18}.

\setlength{\parindent}{0pt}\underline{\textbf{Evaluation metrics.}} For each dataset, performance is evaluated by measuring index construction time against the recall rate with a fixed number of CPU cores, while scalability is assessed by tracking construction time relative to the number of CPU cores, keeping the recall rate constant. To ensure fairness, we tested the proposed method using all graph structures employed by the compared methods and averaged the results to minimize variations. To ensure fairness, we tested the proposed method using all graph structures employed by the compared methods and averaged the results to minimize variations.

\begin{table}[t]
\caption{Statistics of real-world datasets.}
\begin{tabular}{l|l|l|l|l}
\toprule
\textbf{Dataset} & \textbf{Dim} & \textbf{\# Base} & \textbf{\# Query} & \textbf{LID \cite{amsaleg2015estimating}} \\ \hline
SIFT1M\cite{anon2010}            & 128            & 1,000,000            & 10,000            & 9.3            \\ \hline
SIFT1B\cite{anon2010}            & 128            & 1,000,000,000            & 10,000            & 12.9           \\ \hline
GloVe\cite{pennington2015glove}         & 100           & 1,183,514           & 10,000           & 20.0           \\ \hline
ISD3B           & 256           & 3,645,232,672           & 10,000           & 29.1           \\ \hline
VDD10B           & 512           & 10,483,835,016           & 10,000           & 10.9           \\ \bottomrule
\end{tabular}
\label{tab:datasets}
\end{table}

\subsection{Experimental Result and Analysis}

\subsubsection{\textbf{Construction Performance}}
As shown in Figure~\ref{Figure 3} (left column), \textbf{SOGAIC} achieves an average time reduction of 47.3\% across all datasets larger than \textbf{\textit{SIFT1M}} compared to the baselines. For datasets exceeding one billion points, such as \textbf{\textit{SIFT1B}}, the HNSW implementation in Faiss failed to complete due to out-of-memory issues. On the \textbf{\textit{ISD3B}} dataset, which has a high LID value similar to \textbf{\textit{GloVe}}, DiskANN failed to complete due to severe data partition imbalance, while SPTAG required significantly more time to achieve higher recall, as it performs numerous iterations to introduce sufficient overlap. In contrast, \textbf{SOGAIC} not only resolved the overload issue but also reduced the overlap needed to maintain comparable accuracy, decreasing it from the preset maximum overlapping factor ($\boldsymbol{\Omega}$=4) to an average of 1.93 subsets per vector (total points to build for all subsets / base points). This 51.8\% improvement is achieved through the use of a finely tuned adaptive parameter ($\epsilon$=1.8), which dynamically adjusts the assignment process to balance efficiency and accuracy without unnecessary computational overhead.

\subsubsection{\textbf{System Scalability}}
\begin{figure}[t]
  \includegraphics[width=1\columnwidth]{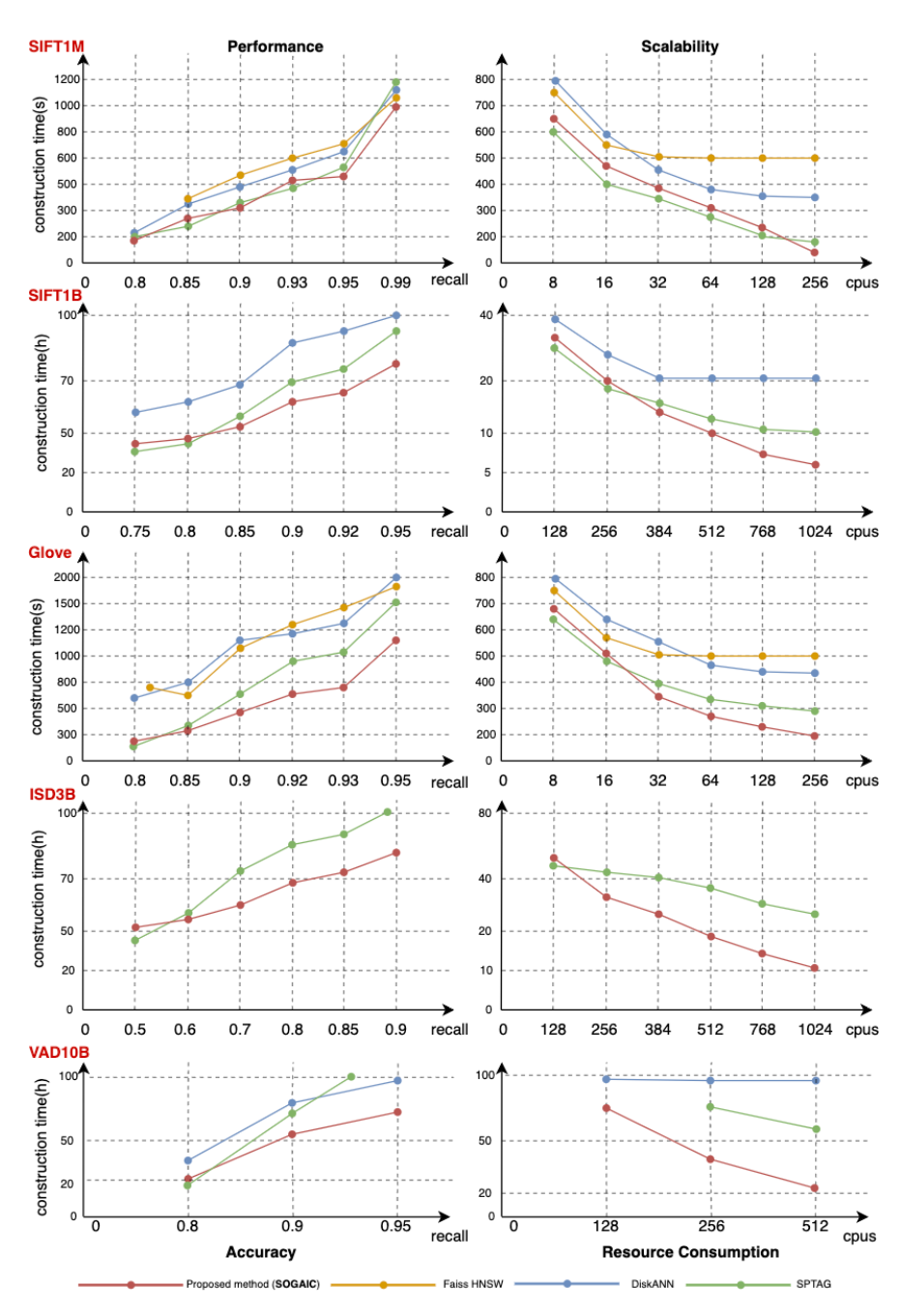}
  \caption{Comparisons of the proposed method (SOGAIC) with others (Faiss HNSW, DiskANN, SPTAG) on both performance and scalability for different datasets: \textbf{\textit{SIFT1M}}, 
  \textbf{\textit{SIFT1B}}, 
  \textbf{\textit{GloVe}},
  \textbf{\textit{ISD3B}},  
  \textbf{\textit{VDD10B}}.}
  \Description{Experiment result. }
  \label{Figure 3}
\end{figure}

As shown in Figure~\ref{Figure 3} (right column), the results demonstrate that the proposed system, \textbf{SOGAIC}, significantly reduces construction time as more computational resources are utilized. Unlike other baselines, \textbf{SOGAIC} maintains a near-linear relationship between performance and resource usage across all datasets, making it highly scalable with the addition of machines. For the ultra-large-scale, high-dimensional dataset \textbf{\textit{VDD10B}}, SPTAG failed to complete within 100 hours when attempting to achieve a previously controlled recall rate (0.95) with 128 CPUs. Meanwhile, DiskANN suffered from consistently high construction times across all resource configurations due to poor scalability. In contrast, \textbf{SOGAIC} completed the process in approximately 80 hours, further reducing the time to under one day with 512 CPUs. This improvement is attributed to the load-balanced scheduling framework and the agglomerative merging strategy.

\section{Conclusion}

In this paper, we introduce \textbf{SOGAIC}, a scalable and overload-aware system specifically designed for constructing ultra-large-scale graph-based ANNS indexes. To address the performance and scalability limitations of existing systems, we propose a dynamic data partitioning algorithm that adaptively introduces overlaps among subsets while adhering to overload constraints, along with a distributed load-balancing subgraph construction framework with an agglomerative merging strategy. Experimental results demonstrate that \textbf{SOGAIC} delivers a 47.3\% average performance improvement compared to existing methods while preserving scalability. Furthermore, this approach has been successfully deployed in a real-world industrial search engine, managing over 10 billion daily updated vectors and serving hundreds of millions of users.

\balance

\end{document}